\documentclass[10pt]{article}

\usepackage{cite} 

\usepackage{hyperref}

\usepackage{graphicx}
\usepackage[font=scriptsize]{caption}

\usepackage[a4paper,top=2cm,bottom=2cm,left=3cm,right=3cm]{geometry} 

\begin{document}

\title{\textbf{AstroGarden of Roma Tre University: from presence to online tour}\vspace{-4ex}}
\date{}
\maketitle
\begin{center}

\author{ Adriana Postiglione$^{1,2}$\\
\textit{on behalf of}\\
Ilaria De Angelis$^{1,2}$, Massimiliano Di Blasi$^{1,2}$,}
\end{center}

\paragraph{} \parbox[t]{1\columnwidth}{$^1$Dipartimento di Matematica e Fisica, Universit\`a degli Studi Roma \\Tre, Rome (Italy)\\%
    $^2$INFN Sezione di Roma Tre, Rome, (Italy)\\
    
    adriana.postiglione@uniroma3.it}

\begin{abstract}
The transition of teaching activities to online mode, forced by the Covid-19 emergency, had also positive aspects, as it pushed to create new contents and use new approaches. An example is represented by our experience at the Department of Mathematics and Physics of Roma Tre University, where we had to revolutionize an activity we carried on countless times over the years: the guided visit to our astronomical garden, the AstroGarden. In this paper, we analyze the new approach we used especially regarding the activities with the so-called oriented globe, the different audiences we reached and the positive feedback we received. 

\end{abstract}

\section{Introduction}
Among the scientific topics for which misconceptions seem to be more widespread, those related to astronomy certainly stand out. Over the years, countless studies have shown that people carry wrong or confusing ideas especially when it comes to the relationship between the Sun, Earth and Moon\cite{Sadler, Ault, Schoon1, Schoon2, NASA}. The shape of the Earth and the meaning of latitude and longitude, the phases of the moon, the meaning of seasons, the sense of time zones, the existence of the Tropics: all these topics demonstrated to be the source of numerous misconceptions that prove to be particularly resistant, so much so that they have been detected regardless of age and level of education all over the world \cite{Sneider, Atwood, Baxter, Stover, Zeilik}.

Such diffusion and persistence of misconceptions in this field may be due to the tendency of treating these topics in a purely theoretical way. Hands-on and experimental activities, on the other hand, have proven to be effective in helping people - both in formal and informal learning contexts - to truly understand scientific concepts and to overcome their preconceptions, especially when combined with an explicit treatment of the most common misconceptions \cite{Freeman, Prince, Hake, Oppenheimer, Stover, Chun, Postiglione1, Bishop}.

With this awareness in mind, starting from 2009 at Roma Tre University the astronomical garden ``AstroGarden" was born \cite{Altamore1}, with the intention of creating a place where visitors could firsthand experience celestial phenomena and in particular the relationship between the Sun, Earth and Moon through telescopes and a special globe at the center of the garden, often called oriented globe or self-centered globe, which reproduces the Earth's orientation in space. Over the years, many people have visited this garden: general public on the occasion of the European Researchers' Night or other public events, schools of all levels, pre-service and in-service teachers, kindergarten children.

Recently, due to the Covid-19 pandemic, face-to-face visits have been suspended. This situation forced us to innovate our didactic proposal and create new contents. Although challenging, this condition allowed us to reach more people than before. In particular, we decided to create materials that could be used remotely and autonomously by anyone. The ideal tool for this was the oriented globe, since with the right indications any globe can be arranged to simulate the orientation of the Earth in space, allowing to \mbox{(re-)discover} several celestial phenomena. In this paper we describe the resource we created, the contexts in which we used it and the reactions we received. \\

The remaining paper is organized as follows. In section 2 we describe the main tool we used, the oriented globe, underlying all the topics it allows to address. In section 3 we focus on the material we produced in order to transform the guided visit typically carried out in our AstroGarden into an activity that can also be used autonomously at a later time. In section 4 we analyze the different audiences with whom our material was used, including general public, high school students and pre-service primary school teachers. Finally, in section 5, we sum up our results and outline possible future prospects of our work.

\section{The oriented globe}
The oriented globe, also called self-centered, parallel or day-night globe, is a particular globe able to simulate Earth orientation in space thanks to its peculiar arrangement. Its practice in Italy spread in the 1980s \cite{Lanciano1}, and it has also been used in many ways in different parts of the world \cite{Anati, Bozic1, Bozic3, Bozic3, Shore, Exploratorium, Corbo1, Rossi1, Rossi2}. In 2009, in occasion of the International Year of Astronomy, a fixed version of this globe was built at the Department of Mathematics and Physics of Roma Tre University at the center of its AstroGarden \cite{Altamore2, Postiglione2}(\textbf{Fig. 1}).

Thanks to its peculiar arrangement, this kind of globe shows in real time the pattern of illumination of the Earth’s surface, including its diurnal and seasonal variations. Placing a gnomon, i.e. a small stick of suitable length, on the surface of the globe, it is in fact possible to analyze the length and orientation of its shadow in different places on Earth and thus deduce the corresponding positions of the Sun as hours and days pass (\textbf{Fig. 2}).

\begin{figure}[]
 \centering
\includegraphics[width=0.6\columnwidth]{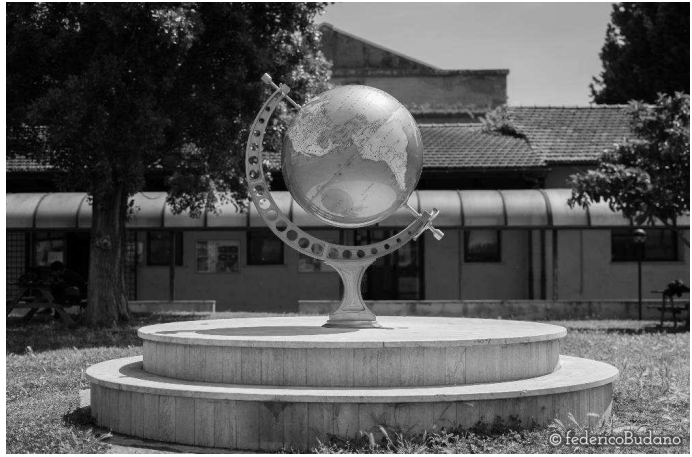} 
\caption{The oriented globe is placed at the center of the astronomical garden of Roma Tre University in Rome. The axis of the globe is inclined with respect to the ground by an angle equal to the latitude of Rome, about 42 degrees. Credits: Federico Budano.}
\label{fig:fig1}
\end{figure}

\begin{figure}[]
  \centering
\includegraphics[width=0.6\columnwidth]{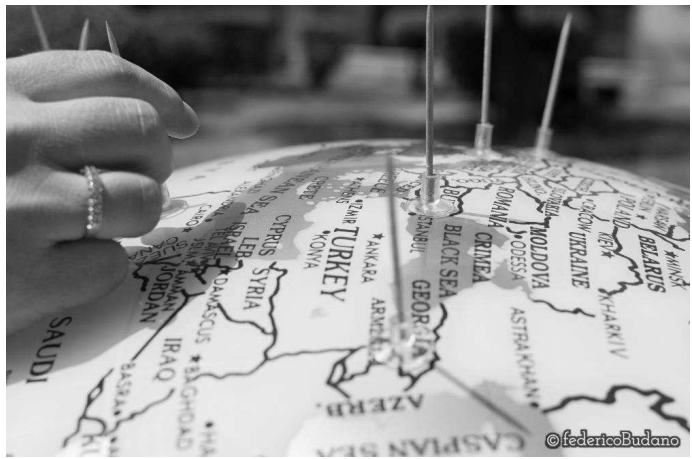} 
\caption{The arrangement of gnomons on the surface of the oriented globe allows to investigate the position of the Sun in different places on Earth as hours and days pass. Credits: Federico Budano.}
\label{fig:fig2}
\end{figure}

Any globe can become an oriented globe, as long as its axis of rotation is parallel to that of the Earth. In order to achieve this, two steps are needed. Firstly, it is necessary to arrange the globe so that its axis of rotation lies in the North-South direction of the place. Then, the city where you are must be placed at the highest point of the globe; this ensures that the inclination of the axis with respect to the horizon coincides with the latitude of the place where the globe is located. For example, if one has Rome at the highest point of the globe, then the rotation axis will be inclined with respect to the horizon by 42 degrees, which corresponds to the latitude of Rome.\\

Once these steps have been performed, the globe is oriented, meaning that, if it is placed under the sunlight, its surface will be illuminated exactly as our planet is at that precise moment. In this configuration, a gnomon placed on the surface of the globe represents a person standing in that place, i.e. their shadows will be parallel and to scale. Since the shadow is directly linked to the position of the Sun in the sky, the gnomon will now allow to investigate time and season of different places on Earth \cite{Corbo2}, giving life to a practical and very effective activity.

\section{The resource developed}
Our aim was to convert the face-to-face visit to our garden into an online visit that could be enjoyed by the audience both in synchronous and in asynchronous mode, i.e. both with the guidance of the University staff and autonomously. In particular, the main target we had in mind was represented by the teachers who typically took their classes to the AstroGarden, who also wanted to feel more confident in replicating our activities on their own.

So we decided to create a video that could simulate the on-site visit through a narrator who guides participants with detailed, step-by-step instructions. In this way, our teachers could either follow the video as a tutorial to prepare their lesson or even show it directly to their students in order to make them discover time-zones and seasons.\\

The video we created, available on YouTube\footnote{The video can be found at: \url{https://www.youtube.com/watch?v=osM8paBQEGE}}, starts with a brief virtual welcome. Then, the oriented globe placed at the center of the garden (\textbf{Fig. 1}) is presented, underlying the position of its rotational axis and the city placed at its highest point.

We thus focus on a common globe, like the ones that can be typically found in schools or easily bought in stores (\textbf{Fig. 3}). We show the steps to follow in order to orient this globe, and we introduce the gnomon as the tool that allows to discover the different phenomena that can be observed with the globe now oriented. 

\begin{figure}[]
  \centering
\includegraphics[width=0.6\columnwidth]{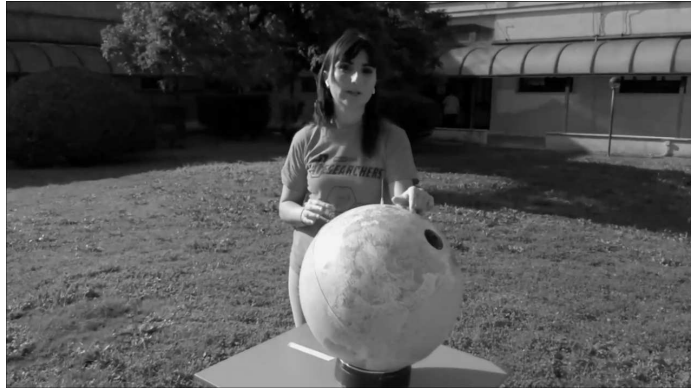} 
\caption{A screenshot of the virtual visit in the AstroGarden, when the orientation and usage of a common globe is described. The full video can be found at: \url{https://www.youtube.com/watch?v=osM8paBQEGE.}}
\label{fig:fig3}
\end{figure}

The gnomon placed on a point of the surface of the globe, in fact, represents a person standing in that location, in the sense that the shadows of the hypothetical person standing up in that point at that moment, and that of the gnomon, are parallel and to scale. Since the shadow projected by a person allows to indirectly derive the position of the Sun in the sky, an oriented globe with gnomons positioned on its surface allows to measure in real time the position of the Sun for every place on Earth carrying out the measurement from a single place, that is the one where the globe is positioned. In other words, the oriented globe allows to virtually travel around the world and discover the corresponding changes in the positions of the Sun.

For example, looking at the direction of the shadow projected on the globe, one can understand if it is morning, noon or afternoon depending on whether the shadow goes west, north/south or east. If you then move from that position along the parallel to the east you are getting closer and closer to the place where the Sun is setting. Vice versa, by moving the gnomon along the same parallel but to the west, you are approaching the place where the Sun is rising. In other words, moving the gnomon along a parallel we are observing not only where day or night is, but also the passing of hours in different places on Earth, and therefore time zones.

When moving the gnomon along a meridian, thus fixing the same time, it is also possible to note that the length of the shadow changes. But even more: it is possible to find a point where there is no shadow at all, that corresponds to the place where the Sun reaches the \textit{zenith} (the ``highest" point of the celestial sphere, the one above the head). If one goes beyond this point, it is possible to see that the direction of shadow is even reversed. This helps to experience the reason why in Rome the Sun can be always found toward the South, while in Buenos Aires always toward the North.

Another important element that can be experienced with the oriented globe is the strange behavior of the Poles. Looking at the zones near the Poles of the globe, in fact, one can immediately notice that they are illuminated in a complementary way at any time of the year and any time of day. Phenomena like the midnight Sun or the fact that at certain latitudes months of darkness alternate with months of light can thus be explained; from this observation, the importance of the Arctic and Antarctic Circles shows up. Moreover, it can be noticed that the North Pole and South Pole experience six months of night and six months of day, as when one is illuminated, the other is in the dark and vice versa, and this remains true even as the hours pass. This leads the way to understanding seasons and experiencing that they are reversed between North and South with respect to the point where the Sun is at its zenith \cite{Corbo2}. 

\section{Contexts in which we used our resource}
The video described in the previous section can be enjoyed autonomously by the audience. However, here we will analyze the use in synchronous mode we have made of it and the feedback we have had in different learning contexts with three different audiences. Specifically, we used it in an informal and engaging mode with the general public and with students, and in a more formal training activity with pre-service school teachers. 

\subsection{\textbf{Engaging mode with public and schools}}

\subsubsection{\textbf{General public}}
We used our video for the first time with the general public on the occasion of the European Researcher’s Night 2020 that took place in full remote mode in November 2020. In particular, we decided to not simply publish it on YouTube as it was, but instead to show it while describing it during a YouTube live broadcast: a University researcher introduced the topics covered and then sent the video on air dividing it into various parts. Thus, between one part and the other of the video, it was possible to interact with the audience and comment and analyze in more detail the various aspects treated with the oriented globe.

Interaction with the audience was possible on two sides. From one hand, participants commented and asked questions via chat, while the researcher answered live by voice. On the other hand, we proposed to participants a series of multiple-choice questions which they could answer directly from their mobile phone, challenging each other to determine the fastest to give the correct answer with the fun and playful \textit{Kahoot!}\footnote{\textit{Kahoot!} is a game-based learning platform: \url{kahoot.com}}\cite{Wang}.\\

Several studies have already demonstrated that \textit{Kahoot!} have positive effects in different learning contexts \cite{Wang, DiBlasi}, and also in our case it was fundamental to engage the audience and make them answer our questions without the fear of being judged, and to keep their attention high. Moreover, the researcher could immediately see the effect of her words in terms of participants’ engagement and comprehension and intervene with further explanations if necessary. The questions we addressed to the public using \textit{Kahoot!} were chosen to analyze the most significant aspects treated with the oriented globe, as well as to bring out and comment on the most common misconceptions, including the significance of seasons and the sense of the astronomical noon.

The broadcast, still available on YouTube\footnote{ The recording of the live streaming can be found at: \url{https://www.youtube.com/watch?v=ULw6iGkZf3o}}, was followed live by 70 participants. Of these, about 40 entered our \textit{Kahoot!} game, with 30 of them answering all questions, while the rest only a few. 

The answers we received to our \textit{Kahoot!} can give us some clues as to the existence of misconceptions among the audience. For example, although 71\% of participants answered that the difference between summer and winter relies on the fact that the Sun reaches a greater height in summer, there is, however, a non-negligible percentage of participants (23\%) who, in any case, showed that the common misconception that in summer the Sun is closer to Earth.

A similar trend can be found about the reason why the poles alternate between months of dark and months of light: 64\% of the participants knew that it was because of the inclination of the Earth rotation axis, while 25\% found the reason in the Earth rotation, and 7\% in the Earth motion around the Sun. Another relevant aspect concerned the position of the Sun at the zenith: 83\% of the answers stated that at noon the Sun can be found at the zenith only in the region between the Tropics; the remaining were divided equally among those who thought that this holds everywhere and those who thought that this holds only at the Poles.

The recording of the broadcast totalled more than 1000 views within a few days. In the weeks after the event, many participants shared with us their satisfaction and enthusiasm about it.

\subsubsection{\textbf{Schools}}
As regards schools, we used the same synchronous mode used for the general public but, instead of YouTube, we used Zoom with individual class groups in order to further encourage interaction by allowing each participant to turn on the microphone and intervene at any time. In fact, two or three students per meeting made comments or questions verbally, for a total of 20 students out of 173 participants of 8 different classes.

In addition to this kind of interaction, we still kept the use of \textit{Kahoot!} and alternated it with further explanations and comments. This ensured a greater engagement of the students, who kept their attention high throughout the lesson in order to correctly answer the questions and rise in the final ranking. It is worth noting that in this case the use of \textit{Kahoot!} allowed us to involve the whole class, making even the most shy students actively participate. In fact, out of 173 participants, 152 students took part in \textit{Kahoot!}.

For this type of audience, about 37\% shows the misconception that the season is due to the distance from the Sun; about 34\% cannot explain correctly why months of darkness and light alternate at the poles, and about 31\% states that the Sun can be at its zenith even outside the region between the Tropics.

\subsubsection{\textbf{Pre-service teachers’ training}}
We also used our video on the oriented globe in synchronous mode as a laboratory proposal for students of the course ``Physics and physics education" of the Primary Education Degree of Roma Tre University. 

The usage of the oriented globe is indeed particularly suitable for primary level of education. The Italian National Indications \cite{ref:Indicazioni}, that delineate the topics to be treat during the school career, cite for primary school pupils the need to be familiar with ``\textit{the periodicity of celestial phenomena (day, night, the path of the Sun in the sky, seasons)}" and the importance of ``\textit{knowing how to orient yourself using the compass and the cardinal points also in relation to the Sun}". Emphasis is also placed for kindergarten children as regards the relevance of ``\textit{developing the first physical organization of the world through concrete activities [...] on the characteristics of light and shadows [...] observing their own movement and that of objects}". 

The oriented globe can thus be used with pupils of various ages with different levels of detail, and it is precisely around this awareness that we built our lesson for aspiring primary teachers. The lesson, held on Teams, lasted two hours; it began with a reflection on the Italian National Indications, and a brief introduction to the meaning of using an oriented globe as a formidable tool for children to first-hand experience important topics such as seasons and time zones. Then, we showed the video in pieces, in order to have the possibility of commenting in depth together on all the aspects treated. Also in this case we kept the use of \textit{Kahoot!} to underline, bring out and analyze the most common misconceptions, in addition to keeping students’ attention high. Here, the use of \textit{Kahoot!} also served to show this platform to pre-service teachers as a tool they could use in their future lessons. 

This lesson, addressed to a maximum number of 100 participants, was held five times in 2021, allowing us to reach 425 students in total.\\

As for the other audiences, also with pre-service teachers we can use \textit{Kahoot!} answers to investigate the spread of misconceptions, in particular as regards seasons and the astronomical noon. 
For example, about the difference between summer and winter, 263 out of 321 students (80\%) were able to relate it to the height of the Sun on the horizon, while 35 of them (10\%) still chose the answer related to the Earth-Sun distance. To the question ``\textit{Why do the poles alternate between months of darkness and months of light?}" 231 students out of a total of 298 (77\%) answered ``\textit{Because of the inclination of the Earth's axis}", while the remaining students divided quite equally among other answers (``\textit{Because of the Earth rotation}" 11\%, ``\textit{Because of the Earth motion around the Sun}" 12\%). Finally, to the question ``\textit{At noon, at what time of the year in Rome is the Sun exactly above our head?}" only 169 participants out of a total of 314 (about 53\%) answered ``\textit{Never}", while a considerable number of people (101 out of 314 - 32\%) answered ``\textit{In summer}".

Although the number of answers to \textit{Kahoot!} does not coincide with the total number of students, all the 425 participants filled in the evaluation questionnaire we administered after the activity. In particular, the lesson was very well received by participants. In fact, 100\% of the students found the lesson interesting and even the 73\% of them found it very interesting. The video demonstrated to be a perfect tool for training teachers: as regards the proposed activities with the oriented globe, 99\% of the students felt that the video and subsequent explanations clearly described the topics covered, and 96\% claimed that they are easy to replicate in school in the proposed modality. In fact, when directly asked if they planned to carry out the proposed activities with their future pupils, 99\% of students answered affirmatively. 

Moreover, several free comments left by the participants stressed once again that they greatly appreciated our didactic proposal, both for the clarity of the resources used to explain the activity with the oriented globe and for the ease of finding the tools necessary to carry out it. In addition, the use of \textit{Kahoot!} has been widely regarded as a precious teaching tool to amuse and keep attention high during online lessons.

\section{Discussion and conclusions}
In this paper we presented the didactic resource we created at the Department of Mathematics and Physics of Roma Tre University to adapt the guided visit to our \mbox{AstroGarden} to the online format. Our resource includes the use of the oriented globe and is aimed at making the audience reflect on different Earth science topics, such as seasons and time zones, thanks to a practical and engaging activity.

The Covid-19 emergency has prompted us to experiment a new online modality that turned out to be very well received by all the audience we addressed: general public, students and pre-service primary teachers. Our proposal represented a good solution during the most acute phases of the emergency, when it was not possible to carry out face-to-face activities for schools and the general public, and for teacher training.

An important feature of our proposal relies on the fact that we managed to guarantee a great interaction with the audience also thanks to the use of \textit{Kahoot!}, which kept the attention high and allowed us to actively involve the great majority of participants, even when it came to high school or university students, who were already worn out by distance learning.

Furthermore, the proposed quiz demonstrated to be a useful tool to probe possible misconceptions present in the audience, in particular as regards seasons and the astronomical noon. Specifically, 20 to 30\% of our audience showed to not being able to correctly explain the meaning of the seasons and the reason why months of darkness and light alternate at the poles; as regards the places of the globe in which the Sun can be found at the zenith, 20 to over 40\% shows to not knowing that these coincide with the regions between the Tropics. We believe that having brought out these misconceptions represents a first step to be able to overcome them.\\

In addition to these considerations, it is important to underline that in the case of the teacher training activity, our resource remains a valuable tool for the future as well. In fact, in creating this resource a lot of attention was dedicated to guide the audience as much as possible in the autonomous reproduction of the activities, so that, both today and in the future, pre-service teachers can view our video at any time in an asynchronous way in order to practice until feeling confident enough to realize the activity - an experiential and hands-on activity - with their future pupils. 
Our experience shows, in fact, that we have reached in an effective way 425 students of the Primary Education Degree Course succeeding in actively involving them in a laboratorial proposal, as indicated by the evaluation questionnaires. In fact, aspiring primary teachers who participated in our lesson greatly appreciated the simplicity of reproducing our activities with the oriented globe as well as our approach in explaining them. This also means that they have somehow overcome their fear of dealing with physics subjects they often consider too difficult to manage. This result, obtained with such a big number of participants, would have been very difficult to achieve in person, especially considering that this laboratory training took place only during a single year, 2021. For this reason, we plan to continue to use our resource even when all the activities will be held in person again.

Given the high satisfaction observed for all the audiences to whom we proposed our activity, its laboratory approach and the fact that it was able to deal with some of the most common misconceptions related to seasons and the astronomical noon, and above of all given the excellent feedback we received concerning its possible realization in the classroom by aspiring primary teachers, we plan to translate the video in English, so that it can be used as a tutorial not only by Italian teachers.

\section*{Acknowledgements}
This work was supported by the Italian Project ``Piano Lauree Scientifiche" and the Young Minds section of Rome of the European Physical Society. Thanks to Federico Budano for the pictures of the oriented globe placed in the AstroGarden. 
The authors also thank the teachers and students of all levels who participated in our visit to the AstroGarden over the years, making it increasingly clear and useful in order to address the topics treated.

\end{document}